# Making the leap to a software platform strategy: Issues and challenges

Yaser Ghanam [a], Frank Maurer [a], Pekka Abrahamsson [b]

[a] Department of Computer Science, University of Calgary, 2500 University Dr. NW, Calgary, AB, Canada T2N 1N4
[b] Department of Computer Science, University of Helsinki, P.O. Box 68, 00014 Helsinki, Finland

**Abstract**

Context: While there are many success stories of achieving high reuse and improved quality using soft- ware platforms, there is a need to investigate the issues and challenges organizations face when transitioning to a software platform strategy.

Objective: This case study provides a comprehensive taxonomy of the challenges faced when a medium- scale organization decided to adopt software platforms. The study also reveals how new trends in software engineering (i.e. agile methods, distributed development, and flat management structures) interplayed with the chosen platform strategy.

Method: We used an ethnographic approach to collect data by spending time at a medium-scale company in Scandinavia. We conducted 16 in-depth interviews with representatives of eight different teams, three of which were working on three separate platforms. The collected data was analyzed using Grounded Theory.

Results: The findings identify four classes of challenges, namely: business challenges, organizational challenges, technical challenges, and people challenges. The article explains how these findings can be used to help researchers and practitioners identify practical solutions and required tool support.

Conclusion: The organization's decision to adopt a software platform strategy introduced a number of challenges. These challenges need to be understood and addressed in order to reap the benefits of reuse. Researchers need to further investigate issues such as supportive organizational structures for platform development, the role of agile methods in software platforms, tool support for testing and continuous integration in the platform context, and reuse recommendation systems.

## 1. Introduction

One of the areas that has contributed to the advancement of the software engineering field is software reuse. Simply put, software reuse is the notion of building software products using artifacts that were used in building other software products [50]. This definition has grown in complexity as the research area expanded. What is to be reused has also changed overtime. Initially, code reuse was the main objective. Nowadays, reuse includes other artifacts such as design documents, use cases, test cases as well as processes and procedures [44]. Some estimates suggest that 60% of the design of all business applications is reusable [53], and only 15% of software code is unique in a given domain or organization [12]. Many advantages of software reuse have been reported in the literature [19,12,16,25], namely: fast delivery of products as less development and testing is required, reduced development and maintenance costs, improved quality of reused artifacts, reduced risks by reusing a previously proved solution, and better project estimates for time and cost.

⇑

One strategic way to achieve reuse is to adopt a software platform approach. We use the term platform to refer to a set of sub-systems and interfaces that form a common infrastructure from which a set of related products (aka. a product family) can be developed [36]. This involves reusing relevant artifacts in the platform, and then a customization process to produce unique products [25].

The decision to adopt software platforms is a strategic one that is taken at the organizational level [37]. While literature is abundant on success stories of adopting a software platform strategy (e.g. [31,49]), there is a need to investigate the issues and challenges that organizations face when making the leap to software platforms.

The goal of this study is to provide an understanding of the issues and challenges that may hinder the adoption of platforms as a reuse strategy. To achieve this goal, we studied an organization that underwent a transition to a platform strategy. The study was conducted in light of the following research questions:

RQ1. What were the general issues and challenges that the organization faced when making the transition to a software platform strategy?
RQ2. What were the specific issues and challenges imposed by recent trends in modern software engineering such as agile methods [47], distributed development, and flat management structures within the context of the new platform?

The main motivation for conducting this study is that the notion of reuse in general has many advantages to offer, but it has not yet been utilized in its full potential due to a number of unsolved issues. As Frakes and Kang [51] suggest, ''reuse research has been ongoing since the late 1960s and domain engineering research since the 1980s. Much has been accomplished, but there is still much to do before the vision of better system building via reuse and domain engineering is completely achieved.'' Also, according to Garlan et al. [7], between 1995 and 2009 (the year of their publication), the nature of software systems has changed dramatically; yet software reuse remains a challenge. Therefore, in this study, we aim to reach a better understanding of the impediments that stand in the way of adopting platforms as a reuse paradigm.

The significance of our investigation is twofold. For one, it offers practitioners a detailed case study of the expected challenges when adopting software platforms and the abovementioned approaches. The resulting understanding allows business and technical leads to have more reasonable expectations, set more realistic goals, and make better informed decisions when planning changes and allocating resources for transitioning to a platform strategy. Secondly, this investigation opens new directions of research by revealing the organizational and engineering problems associated with platform development that need to be tackled to ensure a smoother transition to software platforms. Discussing remedies for the identified challenges is beyond the scope of this article. However, we dedicate a full section in this article to show – by example – how to use the findings to propose remedies and identify needs for tool support.

The remaining of this article will be structured as follows: Section 2 provides an overview of relevant work in the literature. Section 3 provides an elaborate description of our research methodology. Section 4 describes the issues and challenges identified in this research. Section 5 provides a discussion of how the results of this research can be used to find solutions and build tool support if needed. Section 6 talks about the generality of the findings and possible threats to their validity. Section 7 provides a comparison between this work and other work in the literature. And finally, we draw some conclusions in Section 8.

## 2. Literature review

The transition to a software platform strategy in its different forms (e.g. software product lines) has increasingly become a noticeable trend in the software industry. Literature is available on success stories as well as challenging transitions, although the latter is less abundant. For example, Romberg [49] analyzed a number of successful attempts to adopt software platforms. He listed – as examples – video game consoles, IBM Lotus Notes (which has recently been integrated with IBM's Websphere), and the Apache Server. The author mentioned that one challenge common to all these cases was the need to redefine the platform architecture as the demands of the market kept evolving. Also, Hetrick et al. [52] reported on their experience of transitioning to software product lines – which was carried out in an incremental manner in order to avoid the typical up-front adoption barrier. They asserted that the transition required tackling a number of issues including technical issues (e.g. consolidating core assets, quality assurance), and non-technical issues (e.g. team organization, processes). Moreover, a report on Motorola's experience with transitioning to a reuse strategy [45] identified a number of issues that somehow hindered the transition such as political agendas, substantial time investment up-front, and resistance of senior executives and middle managers in the company. Furthermore, Garlan et al. [7] discussed their experience in trying to adopt reuse as an economical strategy to build software systems. The authors stated that they had understood what components they needed, they had engineered these components for reuse, they had had the skills to implement these components, and they had used these components as prescribed. Nonetheless, their transition to the reuse strategy still failed ''miserably'' due to what they called architectural mismatch. The authors also mentioned a number of other challenges such as building trust in the reusable assets, and coping with the evolution of the architecture.

In an attempt to understand common challenges that face organizations when adopting reuse in its different forms, Schmidt [9] outlined a number of challenges that could drive software reuse in general towards failure if not addressed properly. The author put organizational impediments on top of his list and mentioned other types of challenges including economic challenges, administrative challenges, political challenges, and psychological challenges. Furthermore, in the discussion of the status of software reuse research, Frakes and Kang [51] identified a number of unsolved issues such as scalability issues, large up-front investments needed to adopt a centralized reuse strategy, and safety and reliability concerns. Also, Kulandai et al. [46] discussed challenges in developing a software platform from the perspective of a platform architect. They listed a number of – mainly technical – impediments such as design challenges (e.g. segregation of the presentation layer), integration challenges between components developed by multiple teams, and testing challenges. Griss [32] identified a number of factors that need to be taken into consideration in order to enable the success of software platforms such as the business drive, the architecture, the process, and the organization.

In engineering domains other than the software domain, researchers have reported numerous challenges that stand in the way of a smooth transition to platforms. For example, Muffatto and Roveda [38] analyzed three industrial studies in the electromechanical industry and suggested that some key issues need to be addressed to achieve a successful platform practice such as team management, co-location, and standardization.

Compared to research on reuse challenges, a smaller body of research is available on the impact of modern trends in the software industry such as agile methods and flat management structures on reuse. Some work in the literature focused on integrating reuse and agile methods like the work by Hummel and Atkinson [42], Raatikainen et al. [39], and Ghanam and Maurer [55]. Another realm of research focused on distributed development in the context of reuse such as Dhungana et al. [6] and Schmid [27].

Looking at the literature review above, we believe our research as presented in this article fills a number of gaps. First, the reuse issues and challenges available in the literature are mostly scattered in different areas of research (e.g. management, architecture, component-based software). Therefore, for researchers and practitioners who have an interest in this field, populating a comprehensive list of challenges in a single article would be of a great value. Also, to get an idea of how much progress the reuse community has made in the past decade or so, it is imperative that we take a fresh look at reuse and determine whether the issues and challenges reported in older research endeavors still exist. Furthermore, the literature in the field of software reuse is missing a more in-depth analysis of the impact of modern practices in software engineering on reuse. In this article, we highlight practices like agile methods and distributed development, and we show their effect on the reuse process.

Moreover, the general pattern of the research efforts in this area tends to follow a single–angle exploration in which the researchers study the challenges from a technical perspective, a management perspective, or any other interesting perspective. This kind of investigation has its own advantages, but it may result in an incomplete understanding of the observed phenomena. In this article, we try to fill this gap by allowing a number of perspectives to freely emerge from the collected data – which in turn provides a holistic view of the issues and challenges found. That is, in the article, we see how technical challenges and non-technical challenges interplay in the same organization. Also, the literature in the software engineering domain can benefit from an analysis as to how the issues and challenges of reuse found in other domains (e.g. manufacturing) generalize to software engineering. Identifying this overlap will enable researchers and practitioners in the software engineering field to learn and maybe ''reuse'' some of the solutions implemented elsewhere.

At the end of this article (Section 7), we revisit relevant work in the literature in order to conduct an in-depth comparison between the findings of this research and those of other research endeavors.

## 3. Research method

### 3.1. Research context

Our research was conducted in a software company in Scandinavia. To comply with the non-disclosure agreement signed with the company, we use the pseudonym ''Scandin'' to refer to the company thereafter. In its domain, Scandin is considered one of the most influential players in Europe, and it has a significant impact on the market in North America and other parts of the world. The company provides solution packages to individuals as well as corporations and third-party service providers. Scandin has over 800 employees – about half of them work in software development. We describe Scandin as a medium-scale organization (compared to larger organizations in Scandinavia like Nokia). In addition to its headquarters in Scandinavia, the company has four other locations (aka. business units) in other parts of Europe and Asia. The company uses outsourcing for some software projects. Scandin has a flat management structure in the sense that they have cut middle-management layers and provided a less-authoritative organizational structure to ensure the direct involvement of employees in the decision making process. About 8 years ago, Scandin took its first steps to adopt agile software development. Software projects were mainly centered on the development and maintenance of a single solution that required high responsiveness to market needs in order to be able to compete globally. Recently, the company – driven by its new business strategy – decided to build a portfolio of products to target new markets and provide a range of service packages to online customers. For that purpose, the technical strategy was to implement a software reuse approach, where all products in the portfolio are to be built using a common infrastructure consisting of a number of software platforms built in-house, where parts of the platforms are derived from existing products, and other parts are to be built from scratch. That is, the company needed to build platforms on top of which teams across the company should build products and services. For any given product, there is a backend side and a client side. Products in the portfolio have common aspects in both the backend (e.g. licensing, updating) as well as the client (e.g. user interface library). Therefore, platforms were needed on both sides. At the time of our study, the company was in the transitional phase – some parts of the platform have already been built and used while some other parts were still being constructed.

Using platforms as a reuse strategy is a common technique in software product line (SPL) engineering. An SPL is a family of closely related products that share a common set of features, but they are unique in certain aspects [43]. SPLs offer many advantages such as shortened time-to-market, increased product quality and decreased cost [26]. Most SPL engineering approaches are proactive in nature since they perform a two-phase development process [25]. The first phase is domain engineering in which the domain artifacts (i.e. platform) are built. The second phase is application engineering in which the actual building of individual products occur. Managing commonality and variability between different products is conducted using systematic variability management techniques [25]. Such techniques help achieve a number of benefits including defining the sources of variability and the different variants, tracing variability from the model to the code and vice versa, and communicating variability to the different stakeholders [29]. At the time of our study, although Scandin was aiming for a platform strategy, they did not have any variability management practices in place. Secondly, Scandin did not want to follow the two-phase approach. Rather, they opted for a more reactive approach for reasons that will be discussed later in this article.

### 3.2. Data collection and analysis

#### 3.2.1. Data collection

In our research, data was collected using an ethnographic approach [33]. Ethnography is a data collection approach that involves spending time in the field to make first-hand observations. The researcher interacts with the subjects of interest in a natural (as opposed to controlled) setting in order to obtain a holistic view of the context pertaining to the problem under investigation. The rich data collected over the course of the study – including observations, questionnaires and interviews – requires a methodical qualitative approach to analyze [54].

In our research, the study involved the first author conducting 3–4 full-day visits in the company every week, over a period of 6-weeks. During these visits, we adopted non-participant observation by attending presentations, demos, planning meetings and status-update meetings (aka. scrum meetings). Furthermore, in order to get a first-hand impression of the interactions and communication channels, we arranged with the company to stay in close proximity to people of different roles in the organization, namely: senior managers, architects, team leads, and developers. Over the course of the study, we conducted 16 in-depth interviews with individuals of different teams and roles. The interviews lasted between 25 and 72 min each. The interviews were audio-taped and transcribed (around 200 pages of transcripts – using Times, 12 pts).

In the selection process of interviewees, our goal was to get a sample of individuals that covered the different aspects related to our research interest. Namely, we were interested in the following aspects: management (directors influencing platform-related decisions), platform development (teams developing the platforms), and product development (teams building on top of the platform). The initial group of interviewees was selected collaboratively by the researcher and a liaison in the company. During this initial phase, we used snowball sampling to prepare for the second round of interviews. That is, we used the collected data as well as suggestions from the interviewees to guide the selection process of other interviewees. We interviewed representatives of eight different teams, three of which were working on three separate platforms as part of the common infrastructure. The interviews were semi-structured and took various directions based on the interviewee's responses. The role of the interviewee was also vital in determining the direction and focus of the interview (i.e. managers focused on high level issues, whereas team leads and architects focused on technical details). Generally, interviewees were asked questions to describe their role and team responsibilities, how they relate to other teams, what issues or blockers they have been facing when building or using the platforms, and what things they thought were missing but would be beneficial to have. The interviewees were also asked to explain certain aspects of the platform and sometimes to draw diagrams and figures to illustrate their understanding of the overall architecture. The artifacts produced by the interviewees helped the researchers understand the problem and the context better and revealed important issues underlying communication within and across teams. These artifacts also helped in the interpretation of the data collected during the interviews. We were also granted access to documented material communicated among the upper management to obtain a better understanding of the company's vision and strategy. Data from the interviews, the documents, as well as the researcher's observations and diaries (consisting of hundreds of field notes) were all used to complete this study. The data collection phase stopped when we started to get no new insights from new rounds of interviews.

#### 3.2.2. Data analysis

The collected data was analyzed using Grounded Theory [3]. Grounded Theory is a qualitative research method in which generation of a theory occurs by looking into the collected data for patterns and concepts. We started by iterating over the collected data to assign codes, and we refined these codes as more data was coded. This involved renaming, merging, or splitting some codes multiple times. The codes were grouped into larger representative concepts and categories that evolved through multiple iterations by going back and forth between different interviews and the other data sources. The data that was collected and analyzed during the initial phase of the study was used to conduct selective sampling (as opposed to random sampling) when recruiting participants for the interviews that followed. The taxonomy of issues started to saturate after having analyzed about 70% of the data. Having this taxonomy developed, we compared our findings to the existing body of literature in relevant research areas in order to identify similarities and differences.

## 4. Issues and challenges

This study revealed a set of issues and challenges that medium sized, distributed, agile organizations are likely to face when reuse becomes a strategic objective – especially when their context is similar to the context of Scandin as described earlier. The challenges we encountered in the data are captured in the tree shown in Fig. 1 (next page). In our analysis, we kept this tree at a manageable size by merging similar concepts and limiting the depth of the tree to three levels. We classified the challenges under four main categories, namely: business challenges, organizational challenges, technical challenges, and people challenges. In the following subsections, we discuss each of the four categories in more detail.

### 4.1. Business challenges

By ''business'' we refer to the many aspects involved in running a profitable organization including the organization's vision and strategy, sales and marketing, and competition. Our findings show that there are two main issues that can introduce major challenges to introducing a platform strategy, namely: the business strategy, and product-driven platform development.

#### 4.1.1. Business strategy

Scandin's new business strategy to target a new segment of customers in their market had a huge impact on platform development. The services that had been previously provided to products by the common platforms needed to be adjusted in order to accommodate the new scenarios those products were required to support (e.g. by the marketing department). This resulted in considerable reengineering of some existing components. When we asked about the reason why a specific component of the platform was undergoing major reformation, one of the platform architects responded that it was due to:

> "... the new way [Scandin] wants to make business with customers on the retail and OEM level but also with operators..."

| | | |
|---|---|---|
| Business challenges | Business strategy | - |
| | Product-driven platfom development | instability |
| | | dominance of a mainstream product |
| | | competing goals |
| Organizational challenges | Communication | among platform teams |
| | | between platform teams and application teams |
| | | in distributed development |
| | | between business units |
| | Organizational structure | silos |
| | | decision-making |
| | | stakeholder involvement |
| | Agile culture | feature versus component teams |
| | | team autonomy |
| | | business-value thinking |
| | | product ownership thinking |
| | | agility versus stability |
| | Standardization | of documents |
| | | of practices |
| | | of tools and technical solutions |
| | | of acceptance criteria |
| Technical challenges | Commonality and variability | reuse |
| | | variation sources |
| | | cross-cutting concerns |
| | Design complexity | different actors |
| | | requirement of combinations |
| | | requirement of maximizing reuse |
| | Code contribution | Accessibility |
| | | Platform quality |
| | | Platform stability |
| | Technical practices | testing |
| | | continuous integration |
| | | release synchronization |
| People challenges | Resisting change | - |
| | Technical competency | - |
| | Domain knowledge | - |

Fig. 1. Tree of challenges.

Although this issue is not specific to platform-centric development, the experience of Scandin shows that when adopting a platform-centric approach, the amount of rework and testing that needs to be done is usually multiplied because changing a platform component has consequences on all products that rely on that component.

#### 4.1.2. Product-driven platform development

In traditional models of building platforms, a platform-then-product philosophy is dominant as evident in practices like software product line engineering, where there is an emphasis on developing domain artifacts and then application artifacts [25].This succession means that an organization does not start building products until development of the platforms underlying these products has made considerable progress. On the other hand, in Scandin, we noticed that platform development was product-driven in the sense that some platforms were derived from a number of existing products as well as from the requirements of an ongoing project that was considered the first adopter of the platform. In that project, the product that relied on the platform was being developed at the same time as the platform.

As explained by technical leads, product-driven platform development was their strategy to: (a) reduce the conceived risk of lost investments due to over-engineering aspects of the platform that cannot be reused later – either because they turn out to be unnecessary or too complex to reuse and (b) achieve a faster return-on-investment by delivering products to end-customers more quickly than they would have been delivered if a sequential approach had been used. However, our findings show that the latter approach introduces its own risks and challenges, such as:

– Instability: Some components in the platform may not be mature enough when they are used in products which causes products depending on them to be unstable. As one of the managers put it:

> "[Some products]break every second build... the platform is not stable enough in which they are building their architectures."

– The dominance of a mainstream product: If the platform development is driven by one product that is considered a main revenue stream, which is the case in Scandin, then the priorities in the platform development are likely to be coupled tightly to the needs of such a product:

> "... we tightly plan our sprints only based on the [mainstream] project priorities... Now it's about the [mainstream] product but then we know that we need to be able to serve the [other] products later on."

This may cause the platform to become under-engineered – meaning that the components may become too specific to the needs of the mainstream product (e.g. a specific operating system) rendering component reuse challenging across other products. In Scandin, some components – that were supposed to be cross-platform[1] – became specific to the operating system that was required by the mainstream product:

> "... to further develop [the platform] we have to take it to cross-platform and operating-system platforms... That's not there."

Furthermore, focusing too much on the needs of the mainstream product causes other teams who have dependencies on the platform-to-be to become ignored and uninvolved.

> "... because we [platform team] are fully allocated in the [mainstream] project, it is tough to get the time to actually take the other parties into consideration."

In Scandin, this issue had strong effects on some teams who chose to start implementing their own components resulting in redundant code.

– Competing goals: Product teams are pressured by their technical leads to start using the platform as soon as possible (to avoid any redundancy). One of the technical leads explains that:

> "... the roadmap and goal [set for the product teams] would be to [reuse] all the [platforms] that have been built for the [mainstream] project."

On the other hand, other product-specific goals such as fast delivery are pushed by the business leads. Considering the overhead associated with making the transition to the platform (e.g. learning, asking for changes, customizations), some teams in Scandin perceived that it was faster (or less burdening) to use their own artifacts than to reuse somebody else's.

### 4.2. Organizational challenges

A wide range of issues and challenges arise due to the nature of platform development that requires participation and involvement at the organizational level as opposed to the team or business unit level. In the following subsections, we discuss the organizational issues we encountered in the data.

#### 4.2.1. Communication

Platform development introduces more dependencies in the organization than what would normally exist without such a strategy. In Scandin, these dependencies exist between the platform teams themselves, the platform teams and the product teams, and the different business units in the organization. Distributed development exacerbates this communication challenge as will be explained.

– Communication among platform teams: Our findings show that platform teams need to communicate for a number of reasons such as: (1) assigning responsibilities to components, (2) resolving dependencies between components, (3) agreeing on protocols and internal interfaces, (4) synchronizing releases, and (5) arranging for resources that need to be shared. In the case of Scandin, one of the main challenges is to motivate the individual teams to talk to each other beyond formal meetings (if any), where things might have been overlooked or misunderstood, and beyond reading documents (if any) that might be outdated. When this motivation is not there, developers resort to their hunches to resolve a dependency or may integrate with other components in a less than ideal way.

> "The biggest challenge [is] to get people motivated when they have a dependency for outside ... to get the communication started. And even though we have things like scrum of scrums.. but it still does not mean that everything will be brought up there."

Also, in cases, where this communication is not effective, teams may work on overlapping areas of the platform causing redundancy and rework as we observed in company.

[1] The term "cross-platform" in the context of this study means that the implementation is agnostic to the operating system.

– Communication between platform teams and product teams: Teams in Scandin need to communicate at this level because platform teams provide services that are consumed by product teams. For one, product teams need to know how to access and integrate with the platform. Also, product teams provide feedback to platform teams on existing features and report missing ones. As one of the platform developers pointed out:

"... for various reasons there might be a product level feature [requested of the platform teams]. There have been a few [cases] where something is needed [from the platform teams] by [the product teams.]"

Achieving this communication, however, can sometimes be tricky. As we noticed in the company, when an issue arose in product development, some developers found it easier and quicker to find workarounds which might be redundant to what already existed in the platform. This not only caused a lot of rework and redundancy in the code, but it also made testing and maintenance cumbersome in the future. Therefore, for this communication to be effective, product teams need to understand the value of keeping communication channels active at all times (i.e., realize the technical problems associated with redundancy).

– Communication in distributed development: In addition to the inherent challenges of communication in collocated development, distribution of teams over the world introduces further challenges. When some distributed teams in Scandin used tools like instant messaging and shared desktop to hold meetings, the communication did not always serve its purpose:

"... [name of a unit in the company] seems not to have too much problem using this communicator and shared desktop and so forth. Some other units have serious problems with that."

In addition to discussing this issue with individuals in the organization, we also attended an online meeting to have a better understanding of the problem. One of the factors that made this communication a challenge was the different time zones which made arranging meetings more difficult, and sometimes resulted in the meeting time being inconvenient to one party. Other factors included the absence of non-verbal cues such as body gestures and facial expressions especially during screen sharing, and the cultural differences between Scandinavia and other parts of the world, where the language or social protocols were a barrier. For instance, in Scandinavia, where the management structure is mostly flat, a verbal agreement on the phone was sufficient for developers to start executing a plan. On the other hand, in other parts of the world, where authority is very hierarchical, the teams could not execute their plans until they got approval from the relevant line of management in their business unit.

– Communication between business units: Because business units shared common platforms, they needed to communicate. We will talk about this aspect of communication when discussing "silos" in the next subsection.

4.2.2. Organizational structure

In this section, we discuss the impact of how the organization is structured in terms of business units, teams, and management on platform development. We focus on three main issues that we found evident in our data as follows.

– Silos: The analogy to a silo is often used to describe the state of a certain part of the organization that seems to stand alone and not interact enough with the other parts. As we illustrate in this study, silo thinking is a result of an organizational structure, where business units or teams act as independent entities with their own local management and no motivation to adhere to a centralized decision-making body or to share information with other units. In the context we studied, the silo could be a single team or a whole business unit. Our findings show that the silo problem is by far the most serious challenge that faces the organization's transition to a global reuse strategy. Individuals of both management roles and technical roles repeatedly mentioned the term "silo" and complained about the matter almost equally, for instance:

"we have business units... How do they communicate today.. not too well. These silos they don't talk too much [to each other]."

Our data revealed a number of reasons for silo thinking, and a raft of consequences they have on platform development as captured in Table 1 (next page). The reasons mentioned in Table 1 can possibly be traced back to the misalignment between the organizational structure (as described earlier in terms of business units, teams, etc.) and the new platform architecture. As suggested by Sosa et al. [40], such misalignment might have been the reason behind some communication problems as described in the previous section, and this in turn created reasons for silos to appear.

– Decision-making: This issue is partially related to the silos problem, but it encompasses other aspects as well. When the teams and business units have a sense of federalism (which is very evident in our data), making decisions related to the platform becomes a challenge. One case we came across in Scandin involved decisions that needed to be taken on whether to reuse an existing platform or take a different direction such as building an independent variant to satisfy a certain business concern. On the one hand, the corporate had economic reasons to push reuse, but at the corporate level it was often hard to see all the intricate details of the specific business concerns, hence making such decisions challenging. On the other hand, when these decisions were left to the concerned teams and business units, a number of issues caused the decision making process to go astray. For one, the individual units tended to choose the direction, where they saw short-term gains as opposed to thinking about the long-term goals (e.g. sustaining the platform). And there was also the "not-invented-here" mentality that biased business units to develop their own components as opposed to reusing others'.

An attempt by the company to have a centralized decision-making body did not solve this problem. Business units were likely to assume ownership of products and therefore they deemed such decisions an internal matter. As a corporate manager explained:

"And then if [the business unit's decisions] don't get approved [by the centralized body], they kind of tend to think that well.. this is our internal decision..."

Moreover, when the corporate made a decision to invest into building a platform, political challenges arose when trying to kill ongoing projects that might have been redundant to the services provided by the platform. Or even more challenging was the attempt to get a business unit to retire their old systems and migrate to the new platform:

"This is of course an organizational issue to say to somebody that this thing that you have been building for five years is actually going to be discontinued.'
'

Table 1
Reasons and consequences of silo thinking.

| | | |
|---|---|---|
| **Reasons for silo thinking in the organization** | | |
| Resource allocation | Business units would rather allocate their limited resources to meet their local deadlines unless they are forced to participate in a corporate level project | ''And the business units are not forced unless there is a big project that forces them to put their resources aside for this kind of activity.'' |
| Specialization | Specialization in a certain domain makes it difficult to understand the benefits of communicating platform related issues to others (e.g. promoting reusable components) | ''. . . if it's a business-specific platform there is no communication outside of that business unit.'' |
| Lack of motivation | Unless the business unit sees a direct value of sharing a platform at the corporate level, they are not willing to do so | ''They don't have any interest whatsoever in taking this [platform] to corporate level unless they have a cost saving reason.'' |
| Competing targets | Focusing on meeting unit targets and disregarding corporate targets makes platforms too specific to the business unit | ''[the business units] will just build for business target, and when business units disassemble, the assets might be useless.'' |
| Apathy | Some component teams are indifferent about anything outside the locality of their team. The quote is the response of a senior software engineer in a component team when asked who drives the requirements of their component | ''I really haven't spent that much time to really figure out how this thing goes from up to down.'' |
| Evaluation apprehension | The fear of being criticized, supervised or controlled inhibits sharing and communication | ''[the business units] feel unease when they have to come and present their plans to the [corporate].'' |
| **Consequences of silo thinking** | | |
| Missing the big picture | This results in not having a common understanding of the platform architecture which in turn causes other problems such as redundancy and false assumptions | ''they're working in their silos and the changes are so, that only the projects [they] have been working on lately have good common view.'' |
| Redundancy | Business units and component teams run the risk of duplicating an existing component that has been developed somewhere else. Sometimes the duplication is a result of not being aware of assets outside the silo, or not willing to reuse something that was invented somewhere else | ''We also have three systems for that, [and] two systems to update the databases, that's awful.. and this is because of the business unit silos.''''. . . because somebody thinks they can't use it because it doesn't have their business unit label on it. . ." |
| No long-term thinking | The challenge here is to strike a balance between meeting the short-term goals of the business unit and the long-term sustainability goals of the platform | ''Business units have to balance their business drive [with] the long-term sustainable architectural base. There is no decision on that.'' |
| No visibility of reusable assets | Platform assets get buried within the business unit or a certain component team which results in a lot of duplication and missed reuse opportunities | ''we have been digging the assets of the company here for the last year trying to hire and elaborate those platforms, get them on the map...'' |
| False assumptions | Poor communication with other business units or teams results in false assumptions about assets. In this case, an internal decision could have cost the company a fortune | ''Later, [a business unit] wanted to do [a service] and proposed doing a new system... The reason was because mobile protocol cannot work the same as Win protocol... which wasn't true. It was an assumption.'' |
| Platform divergence | A given platform can initially be used by different business units but then internal decisions result in different branches of that platform. After a while, the branches diverge so much causing the platform to become too specific to a certain operating system or product, or even causing the loss of a common underlying model | ''in the common library there are OS adaptations in the code branching which is not too healthy... when they have been building this current architectural base, they have been building it in silos.'' ''there was actually separate business units that worked independently and resourced independently. . . So we have kind of the same base model but there is added [parts]. Many flavors from the same model.'' |

– Stakeholder involvement: Due to the fact that platforms are an enterprise-level concern as opposed to product or team level concerns, it seems to be a challenging matter to get all stakeholders involved from the business side and the technical side. For a medium-scale organization like Scandin with hundreds of engineers and other personnel in sales and marketing, the challenge was to first identify who had a stake in a given platform. That is, who should actually be involved in planning, building or using the platform? This became more difficult when the platform was to serve different business units with various concerns and competing goals. Distributed development and outsourcing were other factors that added to the complexity of this issue.

For example, in order to build a platform for licensing in a unified way across all products in a given portfolio, the company first needed to involve all the parties responsible for the older licensing models, and the parties responsible for merging these models into a single unified licensing component in the platform. This meant getting on-board the technical leads and architects representing products using the older models, products that will use the new model, and products that are specific to certain operating systems. From the business side, solution managers and business analysts were also involved to make sure the technical solution did not affect a business case in a negative way (e.g. affecting a revenue stream or an agreement with a third-party).

One way the company tried to overcome this issue was by holding workshops to allow teams to discuss common issues and understand their different needs from the platform:

> ''we have had several workshops with [business unit name] guys and the [another business unit name] guys and we have mapped all the differences.''

Unfortunately, in some cases involving all stakeholders at once was infeasible due to scalability issues. In such cases, the company chose to postpone the involvement of some parties to a later stage:

> ''we have [team's name], I don't think [they are] going to be [involved]in the project... at least not right now.''

#### 4.2.3. Agile culture

During the past decade, agile software development [47] has gained great momentum and found its way to an overwhelming number of organizations of different scales [48]. Agile methods preach a raft of principles and provide a wide range of practices to achieve these principles. Although the initial focus of agile methods was centered on the efficiency of the team as a unit of operation, recently there has been a movement towards scaling agile methods up to the enterprise level [8]. In the data, we found that there is a need to adjust agile principles and practices before they can be employed in a software platform context. In this section, we list some of the challenges imposed by the agile culture in the organization. Discussing how agile methods can contribute positively to platform development is beyond the scope of this article.

– Feature versus component teams: Feature teams usually assume end-to-end responsibilities in a given system by orienting their work around features (aka. stories); whereas component teams focus more on delivering a sub-system (aka. component) that interacts with other sub-systems in order to be useful [5]. Due to the focus on delivering tangible value to the end-customer, agile advocates promote the idea of building teams around features rather than components [5]. For some people in Scandin, this idea created the perception that component teams are always disadvantageous. As one of the technical leads denotes:

  "It's been told to me that it's bad to have component teams in the agile world which cannot be end-to-end responsible."

However, in the context of platform development in this company, there seemed to be a need for a combination of both. The interviewee here explains that certain services are so fundamental and expensive that they may require a dedicated component team as opposed to being maintained by a feature team as part of an ongoing project:

  "End-to-end responsibility, very tough to implement... One thing we need to accept as an agile organization is that there are certain services that are too expensive to develop [as part of] the project."

– Team autonomy: The other issue that was evident in the data is high team autonomy. When members of highly autonomous teams stayed together for a long time, those teams gradually turned into silos. This phenomenon often resulted in the consequences of silo thinking as discussed previously. Moreover, in some teams, high autonomy had an impact on decision-making, where the team considered certain issues internal without paying much attention to the consequences of their decisions on the underlying platform. The decision-making aspect has already been discussed in more detail in a previous section.
– Business-value thinking: In agile organizations, there is a strong emphasis on delivering business value [28]. The challenge, however, is that business value is not always immediately visible or may not influence the customer directly as we saw in Scandin. That is, the transition to a platform strategy might provide for many advantages for Scandin as a business, but from an end-customer's perspective, nothing has changed. Our findings show that in an environment, where there is strong business-value thinking, it is a challenge to motivate certain teams and individuals to invest into adopting the platform. In this quote, a lead architect explains why some teams could not see the business value in transitioning to the platform strategy:

  "[The platform strategy] is new for us, but it's not producing any new stuff for the customers... The whole stuff is invisible for them."

– Product ownership thinking: Some interviewees in our study raised the issue that some teams and product owners in different business units had been very protective of their assets and products making the transition to a platform strategy more difficult. This is mainly because teams owning a certain component preferred dealing with that component rather than retire it and maintain a shared component in the platform. A technical executive explains his strategy in dealing with duplicate components:

  "before [platforms] become de facto, it requires killing duplicate systems and preventing them from coming up again."

Another issue was that when it came to some platform components, product ownership was not as explicit and clear as it was in individual products. That is, in many cases it was not clear who owned a component in the platform that was shared across different teams and products.

– Agility versus stability: As described earlier, in Scandin's case, the platform is product-driven which means that some platforms were derived from a number of existing products as well as from the requirements of an ongoing project. We noticed that this notion introduced the challenge of striking a balance between the stability of the platform and the ability to change often and add features. On the one hand, platform stability was key, because many products relied on the platform as a common foundation and therefore it had to be trusted and not changed very often. Especially for some critical components that are relatively expensive to develop and maintain, being part of an ongoing product development, where a certain degree of instability is inevitable imposes higher risks:

  "... One thing we need to accept as an agile organization is that there are certain services that are too expensive to develop [as part of] the project."

On the other hand, it was also important for the company to respond to the need of the products in an agile manner in order to be able to compete in their specific market.

Another issue under this category was raised by some participants. When a specific product requests a change in the platform that involves a cross-cutting concern such as usability, it will be challenging to make a choice from the two possible options. The first is to honor the change request to satisfy the customer at hand (following agility principles), in which case all products relying on that aspect would be affected (i.e. causing instability). The other option is to ignore the request until it proves to be an issue in other products too, but that may come at the expense of the satisfaction of the customer at hand. This challenge, however, was not supported by a specific example, so we consider it more of a concern than an actual problem.

#### 4.2.4. Standardization

Some of the challenges we came across in the data were related to the lack of standardization in the organization which affected communication and made reuse more difficult.

– Standardization of documents: Some documents are circulated among platform teams, and between platform teams and product teams. When the documentation practices were not consistent, individuals were less likely to refer to these documents. As one of the interviewees stated, standardizing the retrieval process of documents plays an important role:

  "if I have to find how this works, I know where to go find the information and everything is in one place."

Other interviewees pointed out that the inconsistency across teams in the format of their documents and the level of details made finding information more difficult.

– Standardization of practices: When different teams and business units contributed to a shared platform, the lack of standard practices such as code conventions and testing practices appeared to have a detrimental effect on collaboration and made reuse difficult. One of the interviewees asserted that the lack of code conventions was one of the reasons it was difficult for his team to reuse others' code:

> ''there is actually nothing really that would be [considered] program wide like code conventions.''

– Standardization of tools and technical solutions: When each team in the organization makes their own decisions on what tools and technical solutions they want to use in a given project, as the case in Scandin, platform development and use seems to become more challenging. For example, developers in Scandin need to deal with a number of version control systems and a wide range of testing and continues integration tools before they could contribute to the platform:

> ''I wouldn't know where to find all of these guys' code... It's still not company-wide that there would be even like a nice recommendation that everybody [should] use SVN not GIT or CVS.''

– Standardization of acceptance criteria: Teams in Scandin often defined a list of criteria that needed to be met before a feature or a task was considered done. We noticed a range of things that were considered in different teams such as: successful compilation, passing regression tests, having a predefined bare minimum amount of test coverage, and updating relevant documents. The fact that these criteria were not standardized across the organization caused some teams to lose confidence and trust when reusing components developed by others or when referring to documents written by different business units. For example, one team that put a significant emphasis on reliability in their engines refrained from using other code that did not adhere to the same quality standard.

### 4.3. Technical challenges

As shown in the previous sections, developing software platforms is a business and organizational problem. But it is also an engineering problem that imposes many technical challenges. The data we have collected shows that many of these challenges are due to the fact that a platform needs to satisfy a range of varying requirements in a certain domain, and that many products rely on the platform as the foundation of their functionality. The major challenges that we have identified under this category include: commonality and variability, architectural complexity, code contribution, and practices.

#### 4.3.1. Commonality and variability

As a reuse strategy, platforms provide a common infrastructure on top of which different products can be built. However, components in the platform need to accommodate possible variants so that customization is possible for different business and technical needs [35]. Managing commonality and variability is not always straightforward, and that is why commonality and variability management is a topic by itself in fields like software product line engineering [23]. We discuss commonality and variability challenges around three axes: reuse, variation sources, and cross-cutting concerns.

– Reuse: Managing reuse in the organization is essential for a successful platform development. In our study, we found that this entails not only finding opportunities for reuse in new products, but also dealing with existing redundancy. One of the main challenges we came across in this regard was to detect redundancies in legacy code. Developers often use ad hoc techniques to reuse code such as copy-and-paste (i.e. code cloning); and research has shown that code clones are difficult to trace and often introduce bugs in the system [57]. In Scandin, a particular problem with clones was that if a critical change was made to the original code, the duplicates did not get the update, and when they did, they had to be maintained separately. One of the platform teams explained the problem as follows:

> ''we kept having these pieces of code that were copied [from our platform] and pasted somewhere else [in different products]... then we optimized [the code] and nobody gets to use [the optimized version] because it wasn't in any common place and there was no process [to trace reuse instances].''

As we noticed, redundancy also resulted from poor communication between teams, which yielded multiple implementations of similar services at times.

After detecting redundancy, the next challenge in managing reuse is actually dealing with redundant solutions. As one of the technical leads explained, the process of retiring redundant components and replacing them with a common foundation requires meticulous care to ensure a smooth and stable transition:

> ''first you need to unify [the solutions]... If we cannot make those [duplicate solutions] coexist, then one of those need to take the whole responsibility, but it means one of those systems continues in production and others are retired and taken to maintenance only.''

After the duplicate solutions have been abstracted into a reusable component in the platform, there is one more challenge of making the new asset visible for future projects. In Scandin, visibility was an issue:

> ''we are not sharing all the code we could, because it wasn't under [business unit name] before this. So I am not sure if they would even know as well what we already would have available.''

– Variation sources: Assets in Scandin's platforms had to deal with multiple dimensions of variability in the product portfolio, which imposed a real challenge. Some variations were due to business needs which required different models for different types of customers:

> ''we have different license models depending on the business case. We have one model that goes into the stores. Then we have the [third-party] model where we actually sell through the [third-party]... Then for corporate, we have a couple of different models.''

Operating systems were another dimension of variation:

> ''...you have Androids, iPod, iPad, Mac, Mobile Win ... if each OS has a different client code, you might have a different backend.. it becomes very tedious to maintain, it becomes a burden.''

In Scandin, variation also occurred due to the concept of combinations of services, where every product team (or sometimes every customer) should be able to package their own combination of services from the platform.

– Cross-cutting concerns: Things that cut across different products that use the platform (e.g. usability in the case of Scandin) become a challenge in scenarios, where a change is needed only in a subset of the products but not all. This may require treating this concern as a new variation point which adds to the complexity of variability in the platform.

#### 4.3.2. Design complexity

This issue has been brought up by lead architects as one of the main technical challenges in platform design. We investigated this issue further by looking at design artifacts to identify the reasons of added complexity, namely:

– Different actors: When variability in the platform was driven by the business trying to target different markets or customer bases, this yielded multiple actors, each with their own needs of the platform.
– The requirement of combinations: Due to the requirement of being able to combine Scandin's components and services to build unique products (aka. suites), ensuring a smooth integration between these components and resolving their dependencies in the different combinations was not straightforward. Therefore, when a software platform strategy was adopted in Scandin, stronger emphasis was given to modularity and clean interface definitions during the design process.
– The requirement of maximizing reuse: In the design of the architecture, architects also needed to consider the requirement of being agnostic to the hardware platform, operating system, and other sources of variation as much as possible. As described by one interviewee:

> “[the components] are not related to any operating system. And we chose them in a way that whatever language on the client side is used there is always the possibility to create clients for the services.”

#### 4.3.3. Code contribution

This became a real challenge when Scandin decided to not completely separate platform development from product development. That is, in the product-driven platform development model that was adopted, both platform teams and product teams needed to contribute to the platform. This was especially the case in situations, where the platform teams could not keep up with the increased number of feature requests by product teams. In the context of our study, the company had adopted an internal open-source model, where product teams could assume the responsibility of building features into certain parts of the platform in order to support their products if they did not want to wait in the queue. Other parts that were considered too risky to be open were kept closed within the platform teams. Some of the challenges associated with the internal open-source model were as follows:

– Retrievability: Depending on the organizational boundaries between business units, component teams and distributed teams, the platform code were less or more difficult to find. Our participants attributed this to poor visibility of the assets, poor communication between teams and business units, and the lack of standardization in source code control solutions.
– Platform quality: Because the quality of the platform could be significantly affected when different teams change different parts of the platform on regular basis, Scandin had put an auditing program in place, where changes were audited by a code guardian before they could take effect. One of the technical leads explains the process:

> “Projects delivering the features [product teams] can go and modify [the platform] as long as the [platform team] audits that and makes the release based on that [audit].”

An issue was raised by some participants regarding this model which is that with a lack of standardization of acceptance criteria, the auditing process might become a bottleneck at certain times.

– Platform stability: Ideally, the impact of any change to the platform ought to be tested against all products and combinations that use the platform before it could be released. As a technical lead in Scandin explained, in order to assign this responsibility to product teams, it required a technical solution that duplicated the build environment of the platform locally on their machines so they could see its impact before submitting it for an audit.

#### 4.3.4. Technical practices

Some technical practices that had been successfully implemented in the previously single-product-centered culture in Scandin did not scale well when the transition was being made to platform-development. Our data revealed some challenges in such areas as testing, automation, continuous integration, and releases.

– Testing: To ensure the stability of the platform in the liberal environment of Scandin with their open-source model, rigorous testing practices were needed. One of the challenges associated with that was to be able to populate the different product instances that had been built on top of a given platform, and test the impact of a certain change set on these instances. When this process was in place, it needed to be highly and efficiently automated in order to be effective:

> “[teams needed to] test the functionality of the [platform] in all supported product contexts. . . they can automatically - before releasing products - repeat testing on all supported products and platforms.”

Another challenge that often arises when testing products that share a common platform is identifying what should be tested in the platform and what should be tested in the separate products [10]. Scandin was not any different in this regard. One of the problems we noticed was the diffusion of responsibility among platform teams and product teams. Some product teams assumed that platform teams should be the ones taking care of testing changes in the platform, while some platform teams made similar assumptions about product teams. When we asked a platform team member about the comprehensiveness of the test suites in a certain engine, he noted:

> “We are partly relying on that common base of code being linked into other engines that are then again tested, and that code [in the platform] is tested in that [reuse] process.”

– Continuous integration: Teams contributing to the platform needed to ensure that changes – in the most part – are agnostic to the operating system or hardware platform. Therefore, a practical build process needed to be setup in such a way so that the changes were automatically propagated to all the different relevant build environments in the organization. This required a lot of sharing and effective communication among teams and business units, and at the time of our study, it was still not achieved.
– Release synchronization: One of the challenges raised by some architects had to do with managing the versions and synchronizing the releases of different components in the platform to ensure a trouble-free integration at the product level. As for different versions, sometime the company needed to maintain older versions of components that were still in use by some of their customers. And at any given point of time, it had to be clear to the maintenance and support teams which versions of component A were supported by which versions of component B. This had to be clear also to product teams to ensure they made the transition to newer versions in time for their new product releases. One of the team lead explains why these vesions existed:

''If we change the logic it shouldn't break the integration. And if we are going to break the integration, then it's a new version here that should be in sync. . . there will be several [versions] to be able to give time to the client . . . to migrate and to take the latest version.''

### 4.4. People challenges

Software engineering is one of the fields, where the human aspect plays an essential role in the success of any practice. As evident in the data we collected, making the transition to a software platform strategy is challenged by a number of factors related to individuals in the organization, namely:

#### 4.4.1. Resisting change

As we saw in Scandin, making the leap to a software platform strategy required major changes in the organizational culture. While some people found it easier to adapt and take the ride, others seemed to struggle for different reasons as reported by our participants such as: not seeing the value of the change, perceiving the change as inconvenient, and having to make adjustments for the new work environment:

''Some people are very set in their ways and these might be reluctant to change the way they work.''

There were also political factors that inhibited adopting changes proposed by others. For example, some business unit managers resisted the decision to join forces with other units to develop a certain component because for them that meant letting go of their own existing components:

''Convincing business unit heads to let go of their own asset which they control fully to one joint module is challenge number one.''

#### 4.4.2. Technical competency

The importance of the technical experience, knowledge and skills that usually play an important role in software teams is exacerbated in the context of platform development. Developers in Scandin needed to be able to cope with the complexity of the platform architecture, write cross-platform code, and contribute sound technical solutions to testing and continuous integration problems. When we asked one of the technical leads about some components that were specific to certain operating systems, he attributed that to missing skills such as writing cross-platform code:

''We don't really have a large experience - at least in this site office - of writing cross-platform code.''

#### 4.4.3. Domain knowledge

We noticed in the case of Scandin that a good understanding of the domain is vital in developing a useful and reusable platform. Without sufficient domain knowledge, engineers could not make decisions as to what was common and what was variable in a given component. This also affected decisions pertaining to the discontinuation of certain components if the impact of such decisions on the customer base was not understood.

## 5. Implications of the findings

The findings of this study stimulate numerous research questions that are too difficult to populate and address within the scope of one article. Nonetheless, in this section we show how the findings can be used by researchers and practitioners to inform decision makers about changes to the development process or/and the organizational structure. We illustrate this using an example of one of the issues mentioned previously. This section also shows – by example – how the findings can inform the design of new tools.

### 5.1. Research questions and practical implications

In this section, we present a discussion of one of the issues mentioned previously; namely, feature teams versus component teams, as an example of how the findings can be used to propose solutions. The goal of this section is not to discuss the solution in elaborate details, but rather to give directions to researchers and practitioners on how to go about using the findings at hand. We breakdown the process into four main activities: determine focal point, find interrelations, identify research questions, and then identify solutions.

#### 5.1.1. Determine focal point

The focal point is the main issue under investigation. In this example, we are interested in investigating the issue of the coexistence of feature teams and component teams to support reuse. The dotted box in Fig. 3 highlights this focal point.

#### 5.1.2. Find interrelations

We find it imperative to point out that providing a list of action items that address each of the identified issues and challenges may not be meaningful if not accompanied with an understanding of how these challenges interplay. Therefore, after specifying the focal point, we specify the other issues and challenges that are directly relevant to the main issue. Such issues need to be taken into consideration in order to be able to resolve the main issue at hand.

In this scenario, the three issues that are directly relevant are connected with a solid line to the focal point as shown in Fig. 2. One issue is the communication that is needed between the teams. Another issue is the reuse framework within which both types of teams operate after having identified commonalities and variations. And the third issue is how each type of teams could contribute to the code base.

Some other issues may be relevant to the focal point but in an indirect manner (i.e. through another issue) such as the one connected with a dotted line in Fig. 2. Indirect relevance implies that addressing such issues can play a supportive role to address the main issue. In this example, in order to ensure a more stable platform, tool support is needed to provide a better testing and continuous integration process.

We can also look at the diagram in Fig. 2 from another perspective to determine the course of action that needs to be taken. Namely, this can be either process and structural changes or tool support.

#### 5.1.3. Identify research questions

By looking at the focal point and considering the interconnections between the focal point and the other relevant issues, we can formulate a number of research questions. In this example, some of the research questions are:

Organizational/Agile Culture: How can we provide a hybrid team structure that makes use of component teams and feature teams?
Organizational/Communication: How can these teams communicate effectively?
Technical/Code Contribution: Who can make code contributions to components and how can this process be moderated?
Technical/Technical Practices: What tools are needed to support the integration and testing of the code contributed by the different teams?

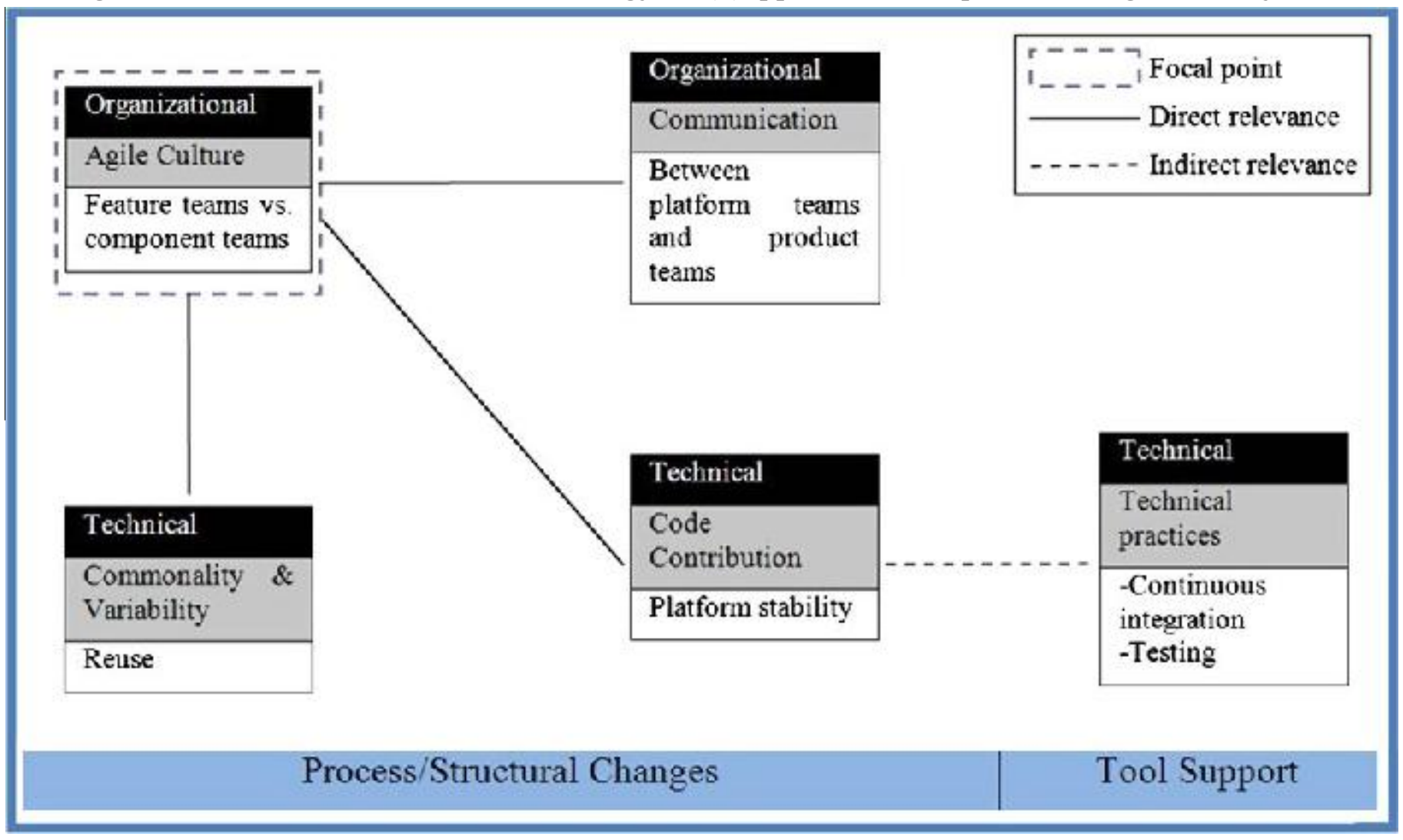

Fig. 2. The main issue of interest and its interrelations with other issues.

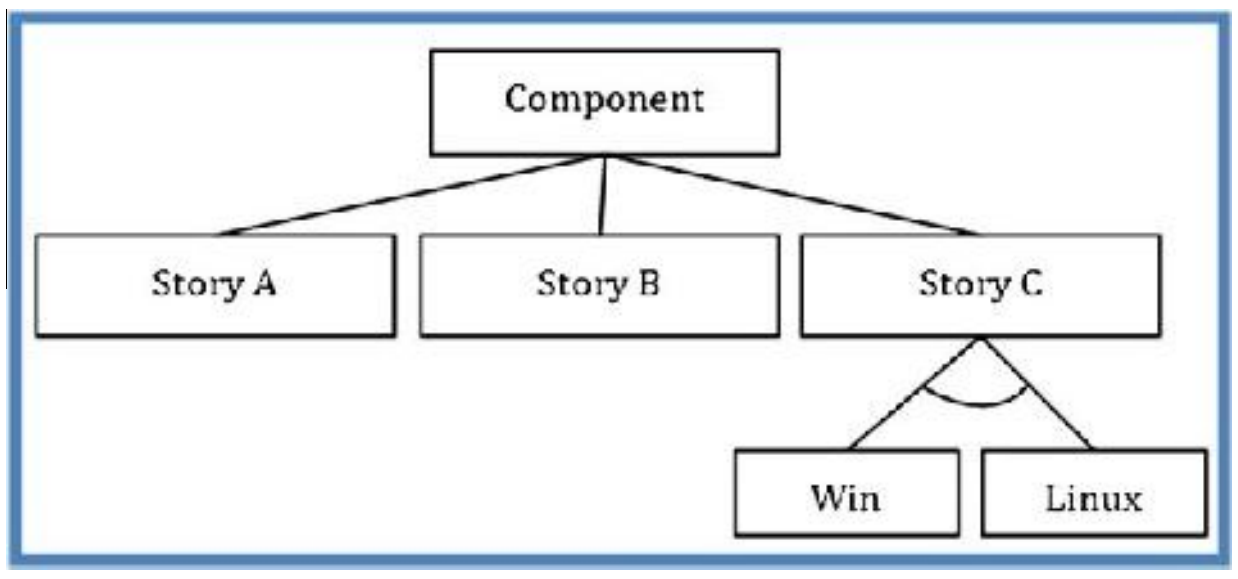

Fig. 3. A feature model to make common and variable aspects more visible.

5.1.4. Identify solution or/and required tool support

The research questions that were determined above will help devise a research methodology and evaluation metric for the proposed solution. Considering the context of the company we have been working with, the following recommendations have been made, and are yet to be implemented:

– A hybrid of feature teams and component teams working together is important. Some component teams are needed to build and maintain certain core services that affect a wide range of products or/and require specific domain knowledge or expertise. Other component teams that have a closer interaction with the application layer may need to collaborate with feature teams.
– Feature teams usually assume end-to-end responsibilities to deliver application level features. To do so, feature teams will use services provided in the platform and may need to modify certain aspects to satisfy their requirements. Modifications can take two forms:
  A. Direct modifications to the core code: feature teams should be able to fill in the missing parts in the platform and possibly add new parts when needed. It is recommended though that this be done under two conditions:
    1. All changes by feature teams need to be audited by the component team owning the part that needs to be changed or extended. This is to ensure that the change is technically sound and does not violate any design or architectural constraints or guidelines. Auditing should also ensure that the code is cross-platform when it needs to be.
    2. Feature teams should have access to the regression tests of the platform components so that they can build locally against different configurations before they commit their changes. Otherwise, the platform may become unstable and may be perceived by its users as unreliable.
  B. Branching from the core code: only if the change that needs to happen is in conflict with other products using the platform and this conflict cannot be resolved should the developers from the component and feature teams collaborate to abstract the common layer and explicate the variable aspects to enable a systematic branching process.
– If a certain requirement from the application level involves a change in the platform, it is beneficial to have at least one member of the component team join temporarily the feature team that is conducting the change. This member will inform the decisions in the feature team before they take place which also makes the auditing process faster. This participation will improve knowledge transfer in the organization because it is an opportunity for developers in the feature team to have a better understanding of the domain and learn about cross-platform issues. In a regular scrum meeting, the member from the component team will then discuss the changes with other members of the component team who may also be reporting about changes from other feature teams (or maybe working on delivering end-to-end features within the scope of their platform).
– Feature teams and component teams need to have a mechanism to communicate how their features interact with other features in the platform. Also, they need to be explicit about variability in the implementation.
– Variability is defined as the capacity of the platform to support a range of products that differ across one or more dimensions. In the case of the company we worked with, the main variability dimension was the operating system. That is, the produced code assets in the platform should be able to support a number of operating systems (e.g. Windows, Mac, Android, etc.). If not supported well, variability may result in some teams not producing cross-platform code, or not testing exhaustively against the different operating systems. Therefore, this dimension of variability should be made more visible by using and communicating a simple model that reflects the common as well as the variable

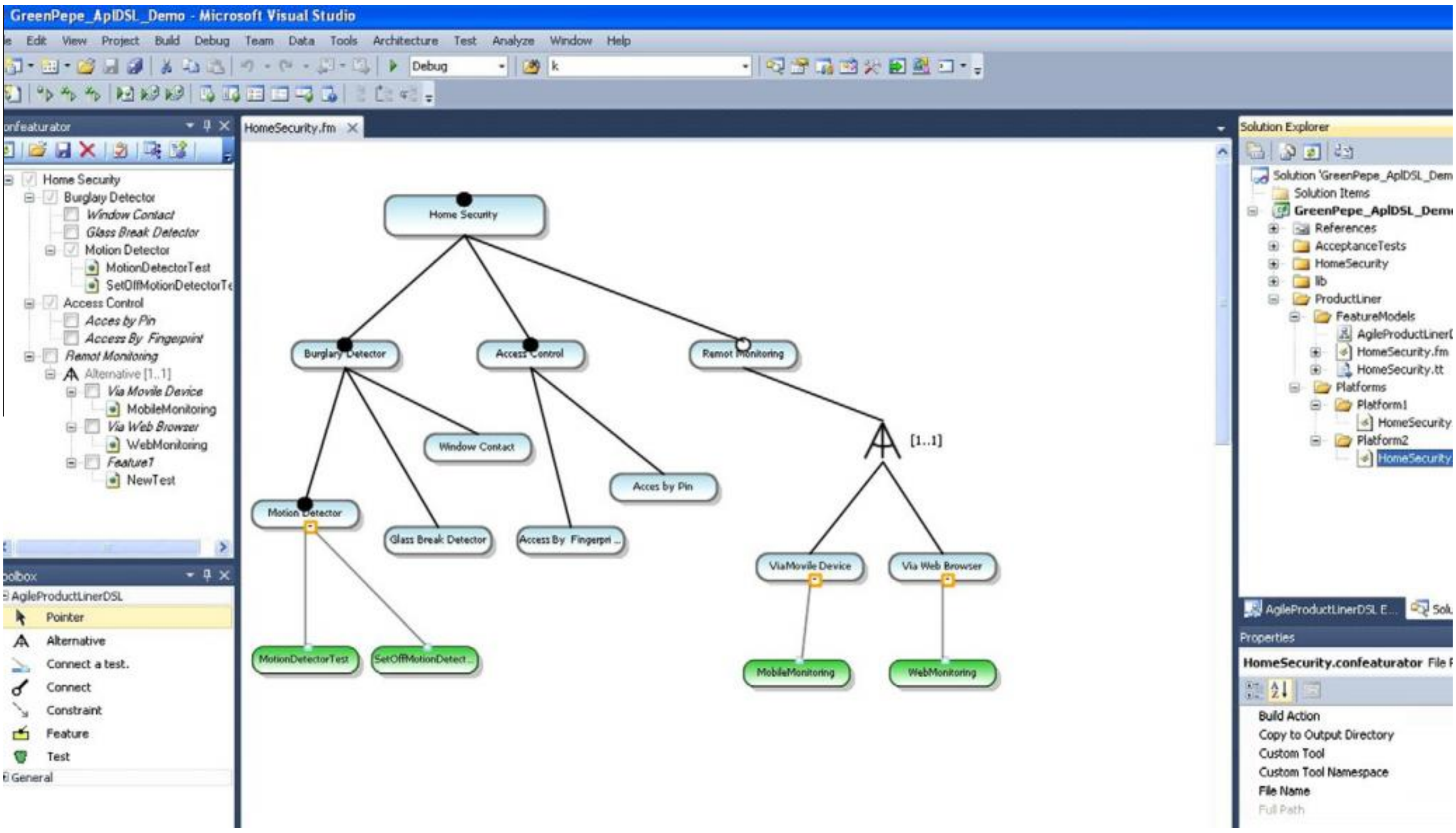

Fig. 4. The tool allows the user to model the common and variable aspects of the platform.

aspects of a given component. Fig. 3 shows an example of how this may be used:

– The above diagram is called a feature model [24]. It can be read as follows: the component consists of three main stories. While Story A and Story B are common across all operating systems (i.e. the implementation needs to be cross-platform), the implementation of Story C may vary based on the operating system (i.e. a different implementation for each operating system). The grouping arch between the Win and Linux options indicates that the options are mutually exclusive. Maintaining such a model increases the visibility of what is common and what is variable in a given component of the platform.
– When feature teams are allowed or even expected to contribute to components in the platform, extra caution should be taken to verify that none of the introduced changes have a cascading negative effect on any products. The following section illustrates how tool support can provide great value in this aspect.

### 5.2. Tool support

Determining the focal point and the relevant issues helps educate the choice of the required tool support and the design of new tools if needed. Although tool support is not always needed to solve a particular issue, in the scenario we describe above, it seemed to be necessary. That is, when feature teams are allowed to contribute to core components, there is a need to make sure that the contributed code does not break other applications that depend on it. Therefore, the build environment required for this context should be capable of ensuring a reliable continuous integration process for multiple products, operating systems, or target machines. This can be done by setting up the build environment so that the builds are done on different target machines automatically and remotely. Current build tools such as Hudson [17] can be extended to work in such a way, but it will require an initial effort to setup.

Moreover, providing direct links between the feature model (as in Fig. 3) and the code artifacts implementing this model is beneficial, because it allows immediate traceability between the different variants (e.g. Win, Android) and the related code artifacts (e.g. classes, methods, etc.). The advantage of doing so is twofold. For one, it enhances maintainability – when a build fails on a Win machine but not on an Android machine, developers will have an easy way to identify those stories specific to the Android machine. If the cause of the failure is not an OS-specific story, then the code that was supposed to be cross-platform is in fact not cross-platform and will need to be modified. Secondly, this traceability provides for easier extensibility because knowledge about how to add support for one more operating system will be accessible to all teams.

To support this notion, we developed a tool called Agile Software Product line Engineering (ASPEN) as an extension to an open source modeling tool available online [13]. ASPEN maintains the links between the model and the code through test artifacts. ASPEN allows the user to model the common and variable aspects of the platform as shown in Fig. 4.

The tool then allows the developers to attach tests to each node in the model (the theoretical foundation of this step and the advantages it provides are explained in detail in [56]). When committing the code changes to the build system, the tool extracts from the provided feature model a profile of tests that need to be run against each operating system. The tool then feeds this information into Hudson that performs a number of builds against different target environments and summarizes the results in a report as shown in Fig. 5. The excerpt of the report in this example shows that the Motion Detector feature is broken on the Windows target machine (highlighted by red). The white cells indicate that a certain feature is irrelevant for a specific target machine. The scope of this article does not allow for an elaborate discussion of the design and implementation of this tool. The goal, however, is to show – through a concrete example – how the findings could be used to determine the required tool support. For more details on this work, we refer the reader to the report available at [14].

| | Android | Windows |
|---|---|---|
| Home Security\Burglary Detector\Motion Detector | | |
| Home Security\Remote Monitoring\Alternative [1..1]\Via Mobile Device | | |
| Home Security\Remote Monitoring\Alternative [1..1]\Via Web Browser | | |

Fig. 5. An excerpt of the report showing the results of executing the features against multiple targets.

## 6. Generality and threats to validity

As in other studies in the literature that rely on data collected from a single company, we cannot claim that the findings of this case study are generalizable to all other companies transitioning to a platform strategy. The study will have to be replicated in a number of other companies in order to confirm generality. In our case study, we considered three different platforms and interviewed individuals of a wide range of roles in the company to get a holistic view of the subject matter and expand the generality of the findings across different roles, teams and technologies.

Moreover, some reported issues may be specific to the cultural context of Scandinavian companies or the domain of the studied company. Another factor that may affect generality is the fact that Scandin did not implement the common domain-then-engineering process to build their platforms, but rather they opted for a more novel approach as discussed previously. In our research, we were not interested in determining whether Scandin made the right choice. Instead, we focused on understanding the obstacles they came across during their transition.

Furthermore, although we assumed that the participants were truthful and honest in their responses and narrations, we noticed that at times the participants were reporting their personal concerns with what might happen as opposed to what was actually happening. We mitigated this issue: (1) by trying to ask for examples and specific incidents whenever an issue was raised, and in the article we make it clear when such examples were not provided and (2) by attending meetings and talking to people from other teams to verify certain claims.

Moreover, the validity of the findings might have been biased by our interpretations. This bias usually comes with all qualitative studies. While qualitative studies typically do not strive for statistical significance, they depend on crosschecking and triangulation of different data sources to verify the findings and draw conclusions. In our case, to mitigate this bias as much as possible, we relied on different sources of data in our analysis such as interview transcriptions, field notes taken during meetings and field visits, and other artifacts provided by the company. We also verified our findings and interpretations with a number of people in the company to check for any misunderstandings.

## 7. Comparison with the literature

In this section, we compare our work with the literature in two steps. First, we discuss the results of our research that are consistent with what is available in the literature. Then, we discuss the new results that add to the existing body of knowledge.

### 7.1. Confirmatory results

In the broader context of the topic, there is a large body of research on platform development in fields other than software engineering. The results of our research confirm many of the findings of such research. For instance, Muffatto [37] analyzed the introduction of a platform strategy in the automobile industry and identified a raft of issues. Some issues are related to the organizational structure, namely: the need for an effective communication structure among platform teams as well as between platform teams and product innovation teams, and the issue of the collocation of platform teams. This goes hand in hand with our findings under the organizational challenges category. Another identified challenge was in regard to the derivation process of platforms from existing products which we also visited in multiple occasions in the article. Moreover, in the manufacturing context, Sundgren [41] points out that the architectural reconfiguration of elements in platform development imposes a real challenge. Our article provided empirical evidence to confirm the existence of similar challenges in the software context.

Moving closer to the software field, Lynex and Layzell [2] identified nontechnical inhibitors of reuse adoption and suggested possible solutions. The authors mention issues similar to the ones we found in our study such as competition amongst business units, unwillingness to share, overlapping responsibilities, quality of components, and other issues. Our findings also support the risks and challenges discussed by Halman et al. [21], namely: integrating existing assets into the platform, the challenge of meeting the needs of all target markets, added complexity in the development process, the need to have a good understanding of the market, and the issue of the flexibility in responding to the market needs versus platform stability. Furthermore, in the context of Hewlett–Packard, Jandourek [11] recognizes teams-structure as a key challenge by affirming that one of the main factors in platform development is an organizational structure that supports interdependencies between platform teams and product teams. The author also addresses concerns similar to ours regarding quality criteria and test procedures, and regarding common development environments and processes which we discussed under standardization. Mili et al. [16] visits the issue of teams-structure in the organization and asserts that a combination of feature and component teams may be necessary.

In the discussion of component-based software engineering, Crnkovic [18] addresses the issue of the sensitivity of platforms to changes. Cusumano and Yoffie [30] offers a thorough discussion on issues similar the ones we found in our study pertaining to cross-platform code such as: synchronizing code bases, keeping track of all variations, and exhaustively testing all versions. Moreover, Greenfield and Short [20] addresses issues such as standardization and automation in production processes. Barnes et al. [4] provided an economic foundation for software reuse in which they mention two source control models: a pure producer–consumer model, and an open source model. In the context of our study, both models were part of the discussion under the code contribution subsection.

Furthermore, at a high level, our classification of issues and challenges provides empirical support to the assertion of Griss [32] that in order for software reuse to succeed, it has to be business-driven, architected, process-oriented, and organized. The unintentional overlap between our work and Griss' is interesting. For one, it underscores the importance of non-technical factors in software reuse compared to technical issues. And secondly, it suggests that over the past decade, the software community has not been satisfactorily successful in finding effective solutions to the obstacles that hinder reuse.

To summarize, in this category of results, our main contribution is seen in confirming the occurrence of the issues and challenges found by other colleagues but in the context of software reuse, and especially during the transitional phase to a software platform strategy. We also found interesting overlap between our findings and the observations made by previous researchers – which clearly indicates that many of the ''traditional'' challenges against software reuse still exist and require further investigation.

### 7.2. New results

This research has exposed a number of points that have not been discussed enough in the literature. From the business perspective, we showed the immediate effect of the business strategy on platform development and how changes in the strategy might hinder progress in building the platform. In our discussion of the organizational structure and decision making, our findings clearly showed that that centralization by itself is not an effective solution, especially in flat organizations. This goes against what was suggested by other work in the literature such as [2] which proposed introducing central coordination of development as a solution to many organizational issues. Moreover, in [21], Halman et al. argue that product families (which are based on platforms) make communication easier. In our findings, on the other hand, communication was found to be a major challenge introduced by platform thinking.

Our discussion of the reasons behind the emergence of silos in a software organization is novel. Furthermore, the thorough understanding of the direct consequences of silos on platform development adds a major contribution to the existing body of knowledge. The article also contributes a new perspective on the role of agile methods in platform development. There has been a number of works in the literature (e.g. [56,15]) that talked about combining agile methods and software product lines (as a platform strategy). But to the best of our knowledge, our work is the first empirical attempt to understand how agile principles and practices might hinder the transition to a software platform strategy if taken as is without serious reconsideration to fit the needs of platform development. For example, we found that the definition of business value, as seen in typical agile circles, may demotivate individuals who work on backend issues that do not have visible effects on the end user. Moreover, the concept of product ownership needs to be revisited in the platform context to define who owns a shared asset, and how different owners should coordinate their needs. Additionally, we discussed team autonomy as another key concept in agile methods, and we showed how it can have a negative impact on the decision making process and might reinforce silo formation.

From a technical perspective, the results of this research may have been touched on by previous work in the literature (e.g. redundancy detection [1], release planning [34]). Nevertheless, our research added more concrete details and specific issues that developers and technical leads encountered in their context. For example, the elaborate description of the internal open-source style and its implications as well as the different issues related to continuous integration and testing provided a solid ground for tool development to support reuse and variability.

In spite of the overlap between our work and the work by Griss' [32] (as discussed in the previous subsection), our work is different in a number of ways. For one, the findings of our research are grounded in a rich set of data that was collected and analyzed following a systematic research approach. Moreover, the lower levels of our categorization provide concrete details about each of the challenges as perceived by the different stakeholders. For example, our discussion of the different organizational and technical issues is considerably richer than the discussion in [32]. Also, our categorization more clearly identifies the human aspect as a key issue in software reuse.

Some of the challenges found in the literature that we did not stumble upon include: finding a balance between the goal of maximizing reuse at one end and the goal of delivering distinctly unique products to the market to drive innovation at the other end [41], customer integration in platform development, and economic justification of platforms [22].

All in all, compared to the existing body of literature, the work presented in this article – to the best of our knowledge – provides a comprehensive list of issues and challenges, both technical and nontechnical. While parts of this research contribute new findings, other parts carry a value in confirming previous findings within the context of software reuse and in elaborating on existing knowledge about certain aspects.

## 8. Conclusion

In this article, we presented an ethnographic case study that aimed to uncover the issues and challenges associated with the transition of a medium-scale organization to a software platform strategy. Throughout the study, we answered the first research question by highlighting the four main categories of issues and challenges. We also tackled the second research question by investigating the impact of the agile culture on this transition, and highlighting the challenges imposed by distributed development. We also discussed how a flat organizational structure affected the decision-making process and imposed challenges related to standardization.

Our findings indicate that the adoption barrier of software platforms in the studied organization was mainly due to non-technical issues including business issues, organizational issues and people issues. There were many technical issues at play as well, but they did not seem to have the same magnitude as non-technical issues. As seen in our findings, organizational issues seemed to have the most visible impact on the smoothness of the transition. One of the most important issues under this category was the emergence of silos. There is an urgent need for further investigation of how silos could be prevented from emerging in the first place. And in case they do emerge, we need to understand how to eliminate them or at least mitigate their negative influence. Furthermore, the organization structure was a key factor in determining the effectiveness of the communication between platform development and product development. The notion that many of the organizational challenges are related to the organization's structure was not particularly surprising given the strong connection that usually exists between the organization's structure and the type of architecture it can support. For examples, issues such as silos and code contribution may exist in any software organization. In platform-centered organizations, however, the special architectural considerations exacerbate the impact of such issues. Therefore, when the transition is being made to software platforms, misalignments between the current organization's structure and the new needs of the architecture need to be identified and addressed.

The results of this research also suggest that with the increasing adoption of agile methods in industry, researchers need to look more into how agile principles can be leveraged to enable reuse and variability. At the same time, researchers should study the detrimental effects some agile principles might have if not tailored to the specific context of software platforms.

Focusing on organizational issues should not result in neglecting technical issues. On the contrary, technical issues should be given their fair share of research, but only in light of the organizational context. Issues related to integration, testing, code redundancy, and code contribution seem to have a strong impact on platform development, especially in iterative and reactive approaches. Tool support is available and it does eliminate some impediments but it is still limited in its ability to provide full support to the different processes in a platform context.

This study provides practitioners with an understanding of the expected challenges surrounding the adoption of software platforms. This understanding will hopefully lead practitioners to make more educated decisions and set more realistic expectations in contexts similar to the one we have analyzed. The study also offers new insights for researchers to further investigate issues such as supportive organizational structures in platform-centered software enterprises, the role of agile methods in software platforms, tool support for testing and continuous integration in the platform context, reuse recommendation systems and many other areas. Future work includes constructing a roadmap for the transition to software platforms in light of the challenges that are well-understood. For other challenges that need further investigation, a root-cause analysis is to be carried out before solutions can be proposed.

## Acknowledgements

We would like to thank the company for allowing us to spend time in their facilities to conduct this research. Also, we would like to extend our gratitude to the individuals who dedicated their time and effort to coordinate or participate in the interviews. Special thanks go to Dr. Jonathan Sillito and the reviewers for their valuable feedback on the content of this article.

This work is part of a broader research funded by Alberta Innovates Technology Futures.